\documentstyle[preprint,aps,eqsecnum,epsfig]{revtex}
\tightenlines
\newcommand{\bea}{\begin{eqnarray}}
\newcommand{\eea}{\end{eqnarray}}
\newcommand{\beq}{\begin{equation}}
\newcommand{\eeq}{\end{equation}}
\newcommand{\bay}{\begin{array}}
\newcommand{\eay}{\end{array}}

\begin{document}
\preprint{\parbox{6cm}{\flushright UCSD/PTH 00-11\\
SLAC-PUB-8446\\ [2mm] May 2000 \\[1cm]}}

\title{Extracting and using photon polarization information in 
radiative $B$ decays}
\footnotetext{Research at SLAC is supported
by the Department of Energy under contract DE-AC03-76SF00515}
\author{Yuval Grossman$^1$ and Dan Pirjol$^2$\\[1cm]}
\address{$^1$Stanford Linear Accelerator Center, Stanford
University, Stanford, CA 94309}
\address{$^2$Department of Physics
University of California at San Diego, La Jolla, CA 92093}
\date{\today}
\maketitle

\begin{abstract}%
We discuss the uses of conversion electron pairs for extracting
photon polarization information in weak radiative $B$ decays.
Both cases of leptons produced through a virtual and 
real photon are considered. Measurements of the angular correlation
between the $(K\pi)$ and $(e^+e^-)$ decay planes in $B\to
K^*(\to K\pi)\gamma^{(*)}(\to e^+e^-)$ decays can be used to 
determine the 
helicity amplitudes in the radiative $B\to K^*\gamma$ decay.
A large right-handed helicity amplitude in $\bar B$ decays is a 
signal of new physics.
The time-dependent CP asymmetry in the $B^0$ decay angular 
correlation is shown to measure $\sin 2\beta$ and $\cos 2\beta$ 
with little hadronic uncertainty.
\end{abstract}

\pacs{pacs1,pacs2,pacs3}

\narrowtext

\section{Introduction}

Rare decays of $B$ mesons have attracted a lot of attention
due to their ability to probe the existence of new physics. The most
accessible such processes are weak radiative decays mediated by
the quark process $b\to s(d)\gamma$. The CLEO Collaboration reported
measurements of both exclusive and inclusive branching ratios for the
$b\to s\gamma$ process, with results in good agreement with the Standard
Model (SM) predictions \cite{CLEO1,CLEO2,CLEOexcl}.
This still leaves open the possibility that new physics could 
be present, but it manifests itself only in the details of the 
decay process, such as polarization effects or differential
distributions. A number of methods have been proposed which 
could detect deviations from the SM predictions along these lines
\cite{Atwood,Mel,Aliev,MaRe,ABHH,KrSe,KKLM}. 

One particular class of methods is based on the SM prediction
that the photons emitted in $b\to s\gamma$ decays are
predominantly left-handed. (Long-distance effects and the light-quark
masses introduce a small right-handed component, which can be neglected
to a first approximation.) This property does not hold true in
extensions of the SM such as the left-right symmetric model 
(LRM) \cite{LRSM},
and therefore, can be used to signal the presence of new physics.
Unfortunately, in $B$ decays all photon polarization information 
contained in the final hadron is lost. 
Since the photon polarization in $\bar B\to K^*\gamma$ decays is 
difficult to detect, most polarization-based methods focused on the 
related $b\to s\ell^+\ell^-$ decay, where the angular distributions and lepton
polarizations can probe the chiral structure of the short-distance
matrix element \cite{Mel,Aliev,ABHH,KrSe,KKLM}. 

Detecting an unambiguous signal for new physics entails a good control
over the SM prediction for the respective exclusive modes. While the
short-distance contribution can be parametrized in terms of (more or
less well-known) hadronic form-factors, the charm- and up-quark
loops introduce hard-to-calculate long-distance contributions.
In $b\to s\ell^+\ell^-$ decays, these effects become significant when $q^2$ 
(the $e^+e^-$ invariant mass) is in the region of the charmonium 
excitations, and are minimal at the lower end ($q^2 \to 0$).
Furthermore, the differential rate is enhanced at this point due to 
the small photon propagator denominator. The remaining long-distance 
contributions, connected with weak annihilation or $W$-exchange 
topologies, can be computed in an expansion in $1/q_0$ \cite{GrPi}.
This suggests that the $b\to s\gamma$ decay, respectively the
$b\to s\ell^+\ell^-$ decay in the small-$q^2$ region, is a good
place to search for new physics through photon polarization effects.

Measuring mixing-induced CP asymmetries in the inclusive $b\to s\gamma$ decay
was proposed as  an indirect method for probing photon polarization 
effects in \cite{Atwood}.
Since both $B$ and $\bar B$ must decay to a common final state,
the resulting asymmetry measures the interference of right- and
left-handed photon amplitudes. As the SM predicts a very small
right-handed admixture of photons in $b\to q\gamma$ decays, a large
mixing-induced CP asymmetry is a signal of new physics.

We explore in this paper an alternative way of measuring the photon
polarization in the exclusive $B\to V\gamma$ decay, which makes use 
of the conversion electron pairs formed by the primary photon.
Electron--positron pairs from photons that were
produced in the inner part of the detector can be traced and their
decay plane can be reconstructed with high accuracy. For example,
at BaBar about 3\% of the photons are expected to convert on the 
beam pipe and the silicon detector \cite{TDR}.
We show in Sec.~II how the conversion process can be used to extract
information about the photon polarization in $B\to V\gamma$ decays, 
in analogy to the classical example of determining the $\pi^0$ parity 
through the chain decay $\pi^0\to \gamma(e^+ e^-)\gamma(e^+ e^-)$ 
\cite{pi0}. The main ingredient of this analysis
is the fact that in $B$ decay into two vectors, the
subsequent decay (or conversion) of the vectors
contains their relative polarization information.
In particular, the angular distribution in the relative angle of the
$K^*\to K\pi$ decay plane and that of the conversion pair can be used
to determine the helicity amplitudes in the $B\to V\gamma$ decay.
In Sec.~III we discuss time-dependent CP asymmetries in the angular 
distribution of the conversion pairs produced in neutral $B$ decays, 
which measure $\sin 2\beta$ and $\cos 2\beta$ 
with little hadronic uncertainty.
We comment in Sec.~IV on the experimental aspects of the methods
proposed here.

\section{Conversion lepton pairs}

There are two possible mechanisms by which the photon (real or 
virtual) emitted
in $b\to s\gamma$ decay can convert into a lepton pair. In the first
one, a virtual photon with momentum $q^2\geq (2m_e)^2$ produces
an electron-positron pair, which are emitted without any other
interaction. In the second mechanism, a real photon produces a lepton
pair which subsequently interacts with a nucleus by Coulomb forces
(the so-called Bethe-Heitler process \cite{BeHe}). 

The lepton pair produced in these two processes are seen very differently
in a detector. The first mechanism produces prompt lepton pairs, which
originate practically from the interaction region. The pairs would be
produced even in vacuum, in the absence of any matter content of the
detector. On the other hand, the Bethe-Heitler process produces lepton 
pairs within the volume of the detector with a probability proportional to
the density of matter.

These arguments can be illustrated with a simple estimate as follows.
The lifetime of a virtual photon contributing to the first mechanism
is (in its rest frame) of the order 
$\tau_0 \simeq 1/\sqrt{q^2} \sim O(1/m_e)$.
In the lab frame the lifetime is longer by a factor of $\gamma =
q_0/\sqrt{q^2}\sim O(m_B/m_e)$.
Thus the photon travels a distance
$\Delta x \sim m_B/m_e^2 \sim 10^{-6}$ $mm$
before decaying. Clearly, any photon that has $q^2 > (2m_e)^2$
travels a distance that is too short to be measured.
For this reason
we will refer to the lepton pairs produced in these two mechanisms as
to short-distance and long-distance conversion leptons, respectively.
In the following we examine them in turn.

\subsection{Short-distance lepton pairs}

In the Standard Model, the decays $\bar B\to X_se^+e^-$ are mediated 
by a
combination of the short-distance Hamiltonian \cite{bsee1,bsee2,bsee3}
\bea
{\cal H}_{\rm s.d.} = \frac{-4G_F}{\sqrt2}V_{tb} V_{ts}^*
\sum_{i=7-10} C_i(\mu) {\cal O}_i(\mu)
\eea
with
\bea
{\cal O}_7 &=& \frac{e}{16\pi^2} \bar s\sigma_{\mu\nu}
(m_b P_R + m_s P_L) b\, F^{\mu\nu}\,,\quad
{\cal O}_8 = \frac{g}{16\pi^2} \bar s\sigma_{\mu\nu}
(m_b P_R + m_s P_L) T^ab\, G^{a\mu\nu}\\
{\cal O}_9 &=& \frac{e^2}{16\pi^2} (\bar s
\gamma_\mu P_L b)(\bar e\gamma^\mu e)\,,\qquad\qquad\quad
{\cal O}_{10} = \frac{e^2}{16\pi^2} (\bar s
\gamma_\mu P_L b)(\bar e\gamma^\mu\gamma_5 e)\nonumber
\eea
and nonlocal contributions introduced by the usual weak nonleptonic
Hamiltonian
\bea\label{Hweak}
{\cal H}_W &=& \frac{4G_F}{\sqrt2} \left\{ V_{ub} V_{us}^*
[ C_1 {\cal O}_1^{(u)}  + C_2 {\cal O}_2^{(u)} ] + V_{cb} V_{cs}^*
[ C_1 {\cal O}_1^{(c)}  + C_2 {\cal O}_2^{(c)} ] \right.\\
& &\left.\qquad \qquad -
V_{tb} V_{ts}^* \sum_{i=3}^{10}C_i(\mu) {\cal O}_i^{(s)} + (s\to d)
\right\}\nonumber
\eea
with 
\bea
{\cal O}_1^{(q)} = (\bar q\gamma_\mu P_L b)
(\bar s\gamma^\mu P_L q)\,,\quad
{\cal O}_2^{(q)} = (\bar s\gamma_\mu P_L b)
(\bar q\gamma^\mu P_L q)\,.
\eea
With new physics other operators can also contribute. In particular, 
the  right handed operators, which can be obtained from the SM operators by 
$R \leftrightarrow L$, are denoted by $\tilde {\cal O}_i$.

The amplitude for $\bar B\to Ve^+ e^-$ decay, with $V$ a light vector meson,
can be written as
\beq\label{M}
{\cal M} = -\frac{4G_F}{\sqrt2} V_{tb} V_{ts}^* e
\left\{
(A_\mu + \frac{1}{q^2} H_\mu) [\bar u(p_{e_+})\gamma^\mu v(p_{e^-})]
+ B_\mu [\bar u(p_{e_+})\gamma^\mu\gamma_5 v(p_{e^-})]\right\}\,,
\eeq
with $A, B, H$ hadronic matrix elements. $B_\mu$ receives only
contributions from the operator ${\cal O}_{10}$ above, and can be expressed
in terms of form-factors alone. $A$ and $H$ can receive contributions from
all the other operators. The part which contains a pole
at $q^2=0$, arising from the propagator of an intermediate photon,
dominates in the small-$q^2$ region, in which we will be
interested in the following. Therefore, we will neglect the nonpole terms
proportional to $A_\mu$ and $B_\mu$.

The matrix element $H_\mu$ is parametrized 
in the most general form in terms of three invariant form factors
\bea\label{Hparam}
H_\mu(q^2) = 
A_\parallel(q^2) \left( q_0 \epsilon_{V\mu}^* - 
v_\mu (q\cdot \epsilon_V^*)\right) + A_0(q^2) v_\mu 
(q\cdot \epsilon_V^*)
+ A_\perp(q^2) i\varepsilon(q,\mu,\epsilon_V^*,v)\,.
\eea
The amplitudes $A_\parallel(q^2)$ (CP-even) and
$A_\perp(q^2)$ (CP-odd) correspond to the virtual
photon polarization being parallel, respectively transverse to 
that of the vector  meson. $A_0(q^2)$ (CP-even) is related to the 
longitudinal polarization.
The values of these form factors at $q^2=0$ describe the coupling 
of a real photon in the weak radiative decay $\bar B\to V\gamma$. 
Thus, in general, $A_0(0)=0$ and in the SM, where the photon is mostly 
left handed, we have $A_\parallel(0) \simeq  -A_\perp(0)$.

An alternative description of the photon coupling is in terms of 
helicity amplitudes, giving the amplitude for the $\bar B$ meson to 
decay into a virtual photon of well-defined helicity. They are related to
the invariant form factors in (\ref{Hparam}) by
\bea
A_\ell(q^2) &=& \sqrt{q^2}\frac{m_B-q_0}{m_V} A_\parallel(q^2) +
\frac{\vec q\,^2 m_B}{m_V\sqrt{q^2}} A_0(q^2)\\
A_{R,L}(q^2) &=& - q_0 A_\parallel(q^2) \mp |\vec q\,|A_\perp(q^2)\,.
\eea
Here $A_{R}$ is the right (left) handed polarization amplitude 
and $A_\ell$ is the longitudinal polarization amplitude which clearly vanish
in the $q^2 = 0$ limit.
Expressed in terms of helicity amplitudes, the rate 
for radiative decay $\bar B\to V\gamma$ is given by
\bea\label{radiative}
\Gamma(\bar B\to V\gamma) = \frac{G_F^2|V_{ts}
V_{tb}^*|^2}{\pi m_B^2}E_\gamma (|A_R(0)|^2 + |A_L(0)|^2)\,,
\eea
where $E_\gamma$ is the photon energy in the $\bar B$ rest frame.

In the SM, the $\bar B\to V\gamma$ amplitude is
dominated by the operator ${\cal O}_7$, which contributes mainly
to $A_L$, with a small right-handed admixture due to long-distance
effects and light quark masses. In certain extensions of the SM, 
such as the left-right symmetric model \cite{LRSM}, a new
penguin operator $\tilde {\cal O}_7$ is introduced involving
right-handed photons.
This operator can make a significant contribution to $A_R$.
Thus, the photon amplitudes are given, in the general case, by
\bea
A_R(0) &=& -(\tilde C_7 m_b + C_7 m_q)
\frac{e}{16\pi^2} 2(m_B^2-m_V^2) g_+(0)
+ a_R\,,\\
A_L(0) &=& (C_7 m_b + \tilde C_7 m_q) 
\frac{e}{16\pi^2} 2(m_B^2-m_V^2) g_+(0) + a_L\,. \nonumber
\eea
where the form factor $g_+(q^2)$ is defined by
\bea
\langle V(p')|\bar q\sigma_{\mu\nu} b|B(p)\rangle &=&
g_+(q^2) \varepsilon_{\mu\nu\lambda\sigma} \epsilon^*_\lambda
(p+p')_\sigma + 
g_-(q^2) \varepsilon_{\mu\nu\lambda\sigma} 
\epsilon^*_\lambda (p-p')_\sigma\\
 & &+
h(q^2) \varepsilon_{\mu\nu\lambda\sigma} (p+p')_\lambda
(p-p')_\sigma(\epsilon^*\cdot p)\nonumber\,.
\eea
$m_q$ is the mass of the strange or down quark depending on the
specific decay.
The long-distance amplitudes $a_{L,R}$ are
introduced by the $b\to c\bar cs$ part of the weak Hamiltonian
and are expected to be about 5\% of the short-distance contribution
\cite{ld1,ld2,ld3,GrPi}. 

We emphasize here that in the SM 
the photon in $B \to V\gamma$ is almost pure left-handed. 
The small right handed
component due to long distance effects and light quark masses is at
the 5\% level. Thus, any measurement of a significant right handed
amplitude will be an unambiguous signal for new physics.

A measurement of the individual 
helicity amplitudes $A_{L,R}(0)$ can therefore give useful 
information about the short-distance weak radiative Hamiltonian.
We will show in the following how a study of the
decay $\bar B\to Ve^+ e^-$ in the low $e^+ e^-$ invariant mass region can 
be useful in this respect.
The argument is a simple adaptation of the classical analysis of Kroll 
and Wada given in \cite{KrWa}.

Let us take the virtual photon to be moving along the $+z$ axis,
and the final state meson $V$ in the $-z$ direction, with momenta
$\vec q$ and $-\vec q$ respectively. The photon converts into a 
$e^+e^-$ pair, with the $e^+$
moving at an azimuthal angle $\theta$ with respect to the $+z$ axis
and a polar angle $\phi$. The vector meson decays into two
pseudoscalars, which will be denoted generically by $K(p_K)$ and
$\pi(p_\pi)$ (corresponding to the interesting case $V=K^*$). The
pion momentum $p_\pi$ is parametrized by the angles $(\psi,0)$ with
respect to the $-z$ direction.
The differential decay rate in these coordinates is given by
\bea\label{diff}
\frac{\mbox{d}\Gamma}{\mbox{d}q^2 \mbox{d}\cos\theta 
\mbox{d}\cos\psi \mbox{d}\phi} &=& {\cal C}
\left\{4|A_\ell|^2 \sin^2\theta \cos^2\psi + 
(|A_R|^2+|A_L|^2) (1+\cos^2\theta) \sin^2\psi\right.\\
& &\hspace*{-3cm}\left.- \sin 2\theta \sin 2\psi \left\{
\cos\phi [\mbox{Re}(A_\ell A_R^*) + \mbox{Re}(A_\ell A_L^*)] +
\sin\phi [\mbox{Im}(A_\ell A_R^*) - \mbox{Im}(A_\ell A_L^*)] 
\right\}\right.\nonumber\\
& &\left. - 2[\mbox{Re}(A_R A_L^*)\cos 2\phi - 
\mbox{Im}(A_R A_L^*)\sin 2\phi]\sin^2\theta \sin^2\psi\right\}
\nonumber\,.
\eea
The constant ${\cal C}$ is given by
\bea
{\cal C} = 
\frac{3\alpha G_F^2 |V_{tb} V_{ts}^*|^2}{8(2\pi)^3 m_B^3}\cdot
\frac{\sqrt{\lambda}}{q^2}\,,\qquad
\lambda = \frac14(m_B^2-q^2-m_V^2)^2 - q^2 m_V^2\,.
\eea

We assumed in deriving (\ref{diff}) that the final leptons are
massless, and neglected the parity-violating effects in the
decay $\gamma^*\to e^+e^-$. Such effects are introduced by 
$Z$ boson exchange (the operator ${\cal O}_{10}$), 
and are parametrized by the hadronic matrix element $B_\mu$ in
(\ref{M}). They 
are negligibly small in the small $q^2$ region
we consider here. The form factor $A_0(q^2)$ vanishes
at $q^2=0$ as $A_0(q^2)\propto q^2$, such that the 
amplitude for producing longitudinally polarized real photons 
$A_\ell(q^2)$ vanishes for $q^2=0$,  as expected.

{}From (\ref{diff}) one obtains, after integrating over $(\theta,\psi)$,
the  following $\phi$ distribution
\bea\label{phi}
\frac{d\Gamma}{d\phi} &=& \frac{32}{9}
\int_{(2m_e)^2}^{q^2_{\rm max}}dq^2 {\cal C}(q^2)
\left\{ (|A_\ell(q^2)|^2+|A_R(q^2)|^2+|A_L(q^2)|^2)\right.\\
 & & \qquad\qquad \left.-
[\mbox{Re}(A_R(q^2) A_L^*(q^2))\cos 2\phi - 
\mbox{Im}(A_R(q^2) A_L^*(q^2))\sin 2\phi]\right\}\,.\nonumber
\eea
In the region close to threshold, the helicity amplitude for
producing longitudinally polarized photons has the asymptotic
form $|A_\ell(q^2)|^2\propto q^2$. Furthermore, to a first approximation,
one can neglect the $q^2$-variation
of the transverse helicity amplitudes $|A_{R,L}(q^2)|$ in this
region and set them equal to their values at $q^2=0$.
Therefore, the $q^2$-integral can be approximated as
\bea\label{phi1}
\frac{d\Gamma}{d\phi} &=& \frac{1}{2\pi}\Gamma(\bar B\to V\gamma)
\left(\frac{\alpha}{3\pi}\log\frac{q_{\rm max}^2}{(2m_e)^2}\right)\\
& &\times
\left\{ 1 - \frac{\mbox{Re}(A_R(0) A_L^*(0))\cos 2\phi - 
\mbox{Im}(A_R(0) A_L^*(0))\sin 2\phi}
{|A_R(0)|^2 + |A_L(0)|^2}\right\} + \cdots\nonumber\,.
\eea
The ellipsis denote a neglected contribution from the longitudinal
amplitude, proportional to the integral
$\int dq^2/q^2 |A_\ell(q^2)|^2$. At $q_{\rm max}^2=1$ GeV$^2$, it 
amounts to about 2\% 
of the leading terms which are kept. Note the presence of the large 
logarithm
$\log \frac{q_{\rm max}^2}{(2m_e)^2}$, which can partly compensate the
additional suppression of $\alpha$ compared to the purely radiative 
rate (\ref{radiative}). Numerically, the value of the suppression 
factor in brackets is $\frac{\alpha}{3\pi}\log\frac{q_{\rm
max}^2}{(2m_e)^2}\simeq
0.01$ at $q_{\rm max}^2=1$ GeV$^2$. This implies an effective
branching ratio of few times $10^{-7}$ for events with $q^2$ in the region of
interest.

{}From (\ref{phi1}) one can see that from the  
$\phi$ dependence in the $q^2$-integrated rate we can extract the ratio
\bea\label{R}
R \equiv \frac{|A_R(0)|\,|A_L(0)|}
{|A_R(0)|^2 + |A_L(0)|^2}\,.
\eea
Combining this with the total exclusive rate (\ref{radiative}),
the individual helicity amplitudes for right- and left-handed photons
can be extracted (up to a $A_R(0) \leftrightarrow A_L(0)$ ambiguity). 
In the SM we expect $R \le 5\%$. Therefore, by measuring $R$
we are sensitive to new physics amplitudes that are an order 
of magnitude smaller than the SM amplitude.

Angular distributions of the type (\ref{phi1}) in the low dilepton 
invariant mass region were also discussed in \cite{Mel,KKLM}.
There, the full expressions for the helicity amplitudes are kept,
including parity-violating effects induced by the operator 
${\cal O}_{10}$. The resulting form of the angular distribution depends
on many form-factors and is therefore not easily connected
to the parameters of the short-distance Hamiltonian. 
In contrast, the phenomenological analysis presented here is 
model-independent;
by restricting to a sufficiently small region above $q^2=0$,
the radiative helicity amplitudes can be directly extracted, without
need for any additional form factors.

\subsection{Long-distance lepton pairs}

Angular correlations in lepton pair production by a real photon
have been suggested long ago as a means for measuring photon 
polarization (for a review see \cite{review}). Our discussion here
will focus on aspects relevant to the photons emitted in exclusive
radiative $B$ decays. 

The cross-section for pair production by a polarized photon was
computed in \cite{BeMa,GHB,GH}. To lowest order it is given
by
\bea\label{sigma12}
\frac{d\sigma}{dE_1 d\Omega_1 d\Omega_2} &=& \frac{Z^2 e^6}{16\pi^3}
\frac{|\vec p_1| |\vec p_2|}{E_\gamma^3 k^4}
\left\{
\frac{(k^2-4E_2^2)(\vec e\cdot \vec p_1)(\vec e\,^*\cdot \vec p_1)}
{(E_1 - |\vec p_1|\cos\theta_1)^2} +
\frac{(k^2-4E_1^2)(\vec e\cdot \vec p_2)(\vec e\,^*\cdot \vec p_2)}
{(E_2 - |\vec p_2|\cos\theta_2)^2}
\right. \nonumber\\
&-&\left.
\frac{k^2+4E_1 E_2}{(E_1 - |\vec p_1|\cos\theta_1)
(E_2 - |\vec p_2|\cos\theta_2)}
[(\vec e\cdot\vec p_1)(\vec e\,^*\cdot \vec p_2) + 
(\vec e\,^*\cdot\vec  p_1)(\vec e\cdot \vec p_2)]\right.\nonumber\\
&+& \left.
E_\gamma^2 \frac{\vec p_1\,^2\sin^2\theta_1 + \vec p_2\,^2 \sin^2\theta_2 
+ 2|\vec p_1| |\vec p_2| \sin\theta_1
\sin\theta_2 \cos(\phi_1-\phi_2)}
{(E_1 - |\vec p_1|\cos\theta_1)(E_2 - |\vec
p_2|\cos\theta_2)}\right\}\,.
\eea
We denote here with $\vec e$ the photon polarization vector and
$p_1 (p_2)$ the positron (electron) momenta.
$k=p_1+p_2-q$ is the momentum transferred to the nucleus,
which will be taken to be infinitely heavy. In this 
limit the nucleus
does not carry away any energy and the photon energy is transferred 
entirely to the electron pair $E_\gamma = E_1 + E_2$. The
angles of the positron and electron with respect to the photon momentum
direction are denoted with $\theta_1$ and $\theta_2$ respectively.

We propose to use as polarization analyzer the rate (\ref{sigma12}) 
integrated
over the electron direction $\Omega_2$. For a linearly polarized photon,
the integrated cross-section for pair production has the following dependence
on the angle $\alpha$ between the photon polarization vector and the projection
of the positron momentum $\vec p_1$ on the plane 
transverse to the photon momentum
\bea\label{sigma1}
\frac{d\sigma}{dE_1 d\Omega_1} =  \sigma_I +
\frac{1}{2}(\sigma_{II} - \sigma_{III}) \cos 2\alpha + \sigma_{IV}
\sin 2\alpha \,.
\eea
We used here the notations of \cite{GH} for the pair production cross-section
by a polarized photon. Only the cross-sections $\sigma_I, \sigma_{II}$
and $\sigma_{III}$ have been computed in this paper (see Eqs.~(17) in \cite{GH})
corresponding to unpolarized photons ($\sigma_I$), and linearly polarized 
photons with the $(\vec p_1\,,\vec q)$ plane parallel to the polarization 
plane ($\sigma_{II}$) and transverse to it ($\sigma_{III}$).
One has $\sigma_I=\frac12(\sigma_{II} + \sigma_{III})$. The cross-section
$\sigma_{IV}$ measures the acoplanarity of the three vectors $\vec q\,,
\vec p_1\,, \vec p_2$. The lepton pair is predominantly produced such that the
photon momentum $\vec q$ lies in the plane of the pair. Therefore, 
$\sigma_{IV}$ is very small and will be neglected in the numerical 
estimates below.

Once the vector meson in $B\to V\gamma$ is observed through its decay
to a pair of pseudoscalars (whose decay plane defines the $x$ axis), 
the polarization vector of the emitted photon is fixed to be 
$\vec e \propto A_\parallel \vec e_1 + iA_\perp \vec e_2$.
The angular distribution of the positron momentum direction is then 
\bea
\frac{d\sigma}{dE_1 d\Omega_1} &=& \Gamma(B\to V\gamma)\sigma_I
\left\{
1 + \cos 2\phi \left[
\frac{\sigma_{II}-\sigma_{III}}{\sigma_I}\cdot
\frac{\mbox{Re }(A_R A_L^*)}{|A_R|^2+|A_L|^2} +
\frac{2\sigma_{IV}}{\sigma_I}\cdot
\frac{\mbox{Im }(A_R A_L^*)}{|A_R|^2+|A_L|^2} \right]\right.\nonumber \\
& &\left.\qquad +
\sin 2\phi \left[
-\frac{\sigma_{II}-\sigma_{III}}{\sigma_I}\cdot
\frac{\mbox{Im }(A_R A_L^*)}{|A_R|^2+|A_L|^2} +
\frac{2\sigma_{IV}}{\sigma_I}\cdot
\frac{\mbox{Re }(A_R A_L^*)}{|A_R|^2+|A_L|^2} \right]
\right\}\,,
\eea
where the angle $\phi$ defined before Eq. (\ref{diff}).
The dependence on $\phi$ has the form 
\beq\label{angdist}
\frac{d\sigma}{d\phi} \propto 1 + \xi R\cos(2\phi + \delta)\,, 
\eeq
with $R$ 
the ratio of amplitudes defined in (\ref{R}), and $\xi$ an efficiency 
factor which can be expressed in terms of the cross-sections 
$\sigma_{I-IV}$ as
\bea\label{xi}
\xi(E_1,\theta_1) = \sqrt{
\left(\frac{\sigma_{II}-\sigma_{III}}{\sigma_I}\right)^2 +
4\left(\frac{\sigma_{IV}}{\sigma_I}\right)^2}\,,\quad
\xi\leq 2\,.
\eea

Most of the pairs produced in the Bethe-Heitler process are emitted
in a small cone of opening angle $2\theta$ with $\theta \simeq 
m/E_\gamma \simeq 0.01^\circ$. We plot in Fig.~1 the efficiency
parameter $\xi(E_1,\theta_1)$ as a function of $\theta_1$ in this 
range, for three values of the positron energy $E_1$. The sensitivity
to the photon polarization is smaller by about a factor of 2 than
for the short-distance lepton pairs (see Eq.~(\ref{phi1})), and is 
maximal when the photon energy is equally distributed among the two 
leptons $(E_1=E_\gamma/2)$.

In Fig.~1 we show also the effective $\xi$ parameter obtained when 
the positron energy is not measured.
This is defined as in (\ref{xi}) but in terms of the cross-sections
integrated over $E_1$. Averaging over $E_1$ further decreases the
sensitivity to the photon polarization. 
An alternative method has been discussed in the literature 
\cite{MaOl} which improves the sensitivity at the cost of statistics.
This method uses as polarization analyzer the rate (\ref{sigma12})
integrated over a restricted region of the electron direction 
$\Omega_2$, 
chosen such that the three vectors $\vec q, \vec p_1, \vec p_2$ 
are almost coplanar $(|\phi_2 - \phi_1 - \pi| \leq \Delta\phi)$. 
A detailed discussion of the resulting asymmetry as a 
function of the width $\Delta\phi$ can be found in \cite{MaOl}.
However, the gain in sensitivity of this
method may be offset by a loss in statistics involved by integrating
over a restricted region in $\Omega_2$.

To conclude this section we stress again the main point. In the 
SM, as $R$ is very small, there is almost no angular dependence 
in the  electron-positron conversion rate.
Any significant measurement of
such angular dependence will be an indication of new physics.
In principle, if indeed such new physics exists, 
using the formulae presented in this section one
could also determine the relative size of this new physics amplitude.
While this may be hard to achieve, the modest goal of
demonstrating any angular distribution may be experimentally feasible
if $R \sim O(1)$.


\section{Time-dependent angular distributions}

A different aspect of polarization
effects in weak radiative decays is manifested through time-dependence
in neutral $B$ decays. Assuming the validity of the 
SM, we will show that certain time-dependent CP asymmetries
involving real or virtual photons can be used to measure 
$\sin 2\beta$ and $\cos 2\beta$ with very little hadronic uncertainty. 
While the measurement of the polarization is sensitive to the
right-handed operator $\tilde {\cal O}_7$, the CP asymmetry is sensitive
to a new CP violating amplitude independent of its
helicity. In the presence of any new contribution to
the decay amplitude with a weak phase that is 
different from the SM phase,
the ``would be'' $\sin 2\beta$ measured in the radiative decay would
not agree with the one measured in $B \to \psi K_S$. Such a 
disagreement will be a clean signal for a new CP violating 
amplitude in $b \to s \gamma$ \cite{GrWo}.
In addition, the sign of $\cos 2 \beta$ can be used to resolve
a discrete ambiguity in the value of $2 \beta$ deduced from the 
measurement of $\sin 2\beta$ \cite{GQ}. Therefore, this measurement
is sensitive to new CP violating contributions to the mixing amplitude.

Before developing the formalism we explain below why we gain 
sensitivity to CP violation phases by using polarization information.
Atwood {\em et al.} \cite{Atwood} studied the 
time-dependent CP asymmetries in $B^0(t)\to V\gamma$
where no polarization information is obtained. They conclude that 
in the SM the asymmetry almost vanishes and only in the 
presence of  right-handed amplitude there is going to be an asymmetry.
The reason for this is simple. In the SM, $\bar B^0$ decays into a 
left-handed photon, while $B^0$ decays into a
right-handed photon. However, interference is necessary in order to 
produce an asymmetry.
Therefore, a final state that is accessible only 
from $B^0$ or $\bar B^0$ does not produce
any asymmetry. In contrast, when we consider decays into a linear 
polarized photon,
both $B^0$ and $\bar B^0$ can decay into the same final state. This
is because the linear polarization state contains an equal mixture
of the circular polarization states.  Moreover, unless there are at 
least two amplitudes with different weak and strong phases,
the magnitudes of the decay amplitudes into a linear polarized state
for both $B^0$ or $\bar B^0$ are the same. 

The situation is very similar to the well known $B^0 \to \psi K$ decay.
The $B^0$ decays only to $\psi K^0$ while the $\bar B^0$ decays into
$\psi \bar K^0$ and there is no interference between the two decays. 
Indeed, if we measure 
$B \to \psi K$ without determining any property of the
final kaon, we do not get any asymmetry. The situation is
very different when we look into final state kaons that 
are admixtures of $K$ and $\bar K$, namely
$K_S$ and $K_L$. In that case, both $B^0$ and 
$\bar B^0$ decay to the same final state and the asymmetry can be used to
measure $\sin 2 \beta$ in the SM.
In both $B \to V \gamma$ and $B \to \psi K_S$ cases
the situation is the same: in order to measure the asymmetry we have to 
observe final states that are accessible from both $B^0$ and $\bar B^0$.   

At this point we already know what can be measured in 
the CP asymmetry in $B \to K^* \gamma$ with linear polarized photon.
Since these final states are
CP eigenstates we can use the well known formalism of 
CP asymmetries in $B$ decays into CP eigenstates.
In the SM (working in the Wolfenstein parametrization)
the $b \to s \gamma$ amplitude has a trivial weak phase, 
and thus the asymmetry is 
sensitive to the mixing amplitude, namely to $2 \beta$. 
Moreover, since we have many amplitudes that interfere we are sensitive to
both $\sin 2 \beta$ and $\cos 2\beta$.
Below we show how to extract $\sin 2 \beta$ and $\cos 2\beta$
from the angular distribution information.

In $B^0\to V\gamma$ decays, final states of well-defined CP
correspond to the amplitudes $A_\parallel(0)$ and $A_\perp(0)$, 
rather than to states of well-defined helicity.
Furthermore, if $V=K^{*0}, \overline{K^{*0}}$, one must
require that the final state be identified through the decay
$K^{*0}\to K_S\pi^0$, which is CP-odd. We denote the corresponding
amplitudes in $\bar B^0\to V\gamma$ decays by $\bar A_\parallel$ and $\bar
A_\perp$\footnote{Note the change in notation compared to Sec.~II.
To conform with usual conventions, $B(\bar B)$ decay amplitudes 
will be denoted with $A(\bar A)$, whereas in Sec.~II we dealt only
with $\bar B$ decay amplitudes.}.

If the decaying meson is tagged as $B^0 (\bar B^0)$ at time $t=0$,
then the amplitudes $A_\parallel$ and $A_\perp$ will depend on $t$
at a later time $t$. This time dependence is given  by 
\bea\label{t1}
A_\parallel(t) &=& A_\parallel(0)\left(f_+(t) + \lambda_\parallel
f_-(t)\right)\\\label{t2}
\bar A_\parallel(t) &=& \frac{p}{q} A_\parallel(0)
\left( f_-(t) + \lambda_\parallel f_+(t)\right)\,,
\eea
and analogous for $A_\perp$ with a different parameter $\lambda_\perp$.
Our notation for the $B-\bar B$ mixing parameters $p$, $q$ is the
standard one \cite{PDG}.
The time-dependence is contained in the functions
\beq
f_\pm(t) = \frac12\left\{e^{-i(m_1-i\Gamma_1/2)t} \pm
e^{-i(m_2-i\Gamma_2/2)t}\right\}\,,
\eeq
with $m_{1,2}$ and $\Gamma_{1,2}$ the masses and widths of the
mass eigenstates of the $B-\bar B$ system.
The parameters $\lambda_\parallel$ and $\lambda_\perp$ are defined as
usual by
\bea
\lambda_\parallel = 
\frac{q}{p}\frac{\bar A_\parallel(0)}{A_\parallel(0)}\,,\qquad
\lambda_\perp = 
\frac{q}{p}\frac{\bar A_\perp(0)}{A_\perp(0)}\,.
\eea

In the following we concentrate on $B^0$ decay via the $b \to s \gamma$ 
quark level transition.
Within the SM, 
the dominance of the left-handed amplitude implies 
\bea\label{SMrel}
A_\parallel\simeq  A_\perp\,, \qquad 
\bar A_\parallel\simeq - \bar A_\perp\,,
\eea
where we used 
$A_L=-A_\parallel + A_\perp \approx 0$ and 
$\bar A_R =  -\bar A_\parallel - \bar A_\perp\approx 0$.
Moreover, in the SM the $b\to s \gamma$ decay amplitude has a
trivial weak phase 
(in the Wolfenstein parametrization) 
and the ratio $q/p=-e^{-2i\beta}$, which gives  
\bea
\lambda_\parallel &=& \frac{q}{p}\frac{\bar A_L+\bar A_R}{A_L + A_R}
\to e^{-2i\beta} \,,\qquad
\lambda_\perp = \frac{q}{p}\frac{\bar A_R-\bar A_L}{A_R-A_L}
\to -e^{-2i\beta}  \,.
\eea

The above results can be generalized to 
many extensions of the SM. When
there is new CP conserving contribution to $A_R$, 
they are not modified.
When there is new CP violating contribution to $A_L$
the above results
still hold, where one must replace $2 \beta$ in both
$\lambda_\parallel$ and $\lambda_\perp$ with the angle between 
the mixing and the decay amplitude. 
When there is new CP violating contribution to $A_R$, but no
strong phase between the left and right handed amplitude, 
$\lambda_\parallel$ and $\lambda_\perp$ are still pure phase. However, the
phase is not the same. Finally, when there is also a strong phase between
the left and right handed amplitudes
$|\lambda|\ne 1$.
While we again assume the SM in the following discussion, 
our results hold also for the first two cases discuss above
(with the general interpretation of $2\beta$).
It is clear how to generalize the results below also to 
more general cases.

Neglecting, as usual, the lifetime difference of the $B_d$ mass 
eigenstates and assuming that $|\lambda| \approx 1$
one finds that each of the time-dependent CP asymmetries for
the final states of linear polarization measure $\sin 2\beta$
\bea
a_\parallel (t) &=& \frac{|A_\parallel (t)|^2-|\bar A_\parallel (t)|^2}
{|A_\parallel (t)|^2+|\bar A_\parallel (t)|^2} = 
-\mbox{Im}\lambda_\parallel \sin(\Delta m t)
\to \sin 2\beta \sin(\Delta m t) ~~\mbox{ (SM)}\\
a_\perp (t) &=& \frac{|A_\perp (t)|^2-|\bar A_\perp (t)|^2}
{|A_\perp (t)|^2+|\bar A_\perp (t)|^2} = 
-\mbox{Im}\lambda_\perp \sin(\Delta m t)
\to -\sin 2\beta \sin(\Delta m t) ~~\mbox{ (SM)}\,,
\eea
whereas the corresponding CP asymmetry in the unpolarized rate is
much suppressed \cite{Atwood}
\bea
A_{CP}(t) &\equiv& \frac{|A_\parallel (t)|^2+|A_\perp (t)|^2-
|\bar A_\parallel (t)|^2 - |\bar A_\perp (t)|^2}
{|A_\parallel (t)|^2+|A_\perp (t)|^2 +
|\bar A_\parallel (t)|^2 + |\bar A_\perp (t)|^2}\\
 &=& 
\frac{|A_\parallel (0)|^2-|A_\perp (0)|^2}
{|A_\parallel (0)|^2+|A_\perp (0)|^2}
\sin(2\beta) \sin(\Delta m t)\,.\nonumber
\eea

It is interesting that much larger asymmetries are obtained 
for the coefficients of
$\sin 2\phi$ and $\cos 2\phi$ in the angular dependence (\ref{phi1}).
Inserting the relations (\ref{t1}), (\ref{t2}) 
into (\ref{phi1}) one finds a particularly simple time-dependence
for the CP asymmetry of the $\cos 2\phi$ coefficient
\bea
\frac{\mbox{Re}(A_R(t) A_L^*(t)) - 
\mbox{Re}(\bar A_R(t) \bar A_L^*(t))}
{|A_R(t)|^2 + |A_L(t)|^2 + |\bar A_R(t)|^2 + |\bar A_L(t)|^2} = 
\frac12 \sin 2\beta \sin(\Delta m t)\,.
\eea
Expressed in terms of the observed time-dependent angular
distributions, this asymmetry can be written as
\bea\label{phit}
4\frac{\langle \cos 2\phi \frac{d\Gamma(t)}{d\phi}\rangle -
\langle \cos 2\phi \frac{d\bar \Gamma(t)}{d\phi}\rangle}
{\langle \frac{d\Gamma(t)}{d\phi}\rangle +
\langle \frac{d\bar \Gamma(t)}{d\phi}\rangle} =
- \sin 2\beta \sin(\Delta m t)\,.
\eea
We denoted here $\langle f(\phi)\rangle = \int_0^{2\pi} d\phi f(\phi)$.
Note that the result (\ref{phit}) does not depend on the 
smallness of the right-handed photon amplitude.
On the other hand, the coefficient of $\sin 2\phi$ is more sensitive 
to the presence of a right-handed photon amplitude
\bea
& &\frac{\mbox{Im}(A_R(t) A_L^*(t)) - 
\mbox{Im}(\bar A_R(t) \bar A_L^*(t))}
{|A_R(t)|^2 + |A_L(t)|^2 + |\bar A_R(t)|^2 + |\bar A_L(t)|^2} = \\
& &\qquad
-\frac{\mbox{Re}(A_\perp A_\parallel^*)}{|A_\perp|^2+|A_\parallel|^2}
\cos 2\beta \sin(\Delta m t) -
\frac{\mbox{Im}(A_\perp A_\parallel^*)}{|A_\perp|^2+|A_\parallel|^2}
\cos (\Delta m t)\,.\nonumber
\eea
Assuming dominance by the left-handed amplitude in the SM 
(see Eq.~(\ref{SMrel})), one
can use this asymmetry to extract $\cos 2\beta$
\bea\label{phitsin}
4\frac{\langle \sin 2\phi \frac{d\Gamma(t)}{d\phi}\rangle -
\langle \sin 2\phi \frac{d\bar \Gamma(t)}{d\phi}\rangle}
{\langle \frac{d\Gamma(t)}{d\phi}\rangle +
\langle \frac{d\bar \Gamma(t)}{d\phi}\rangle} =
-\cos 2\beta \sin(\Delta m t)\,.
\eea
While less clean theoretically than the determination of $\sin 2\beta$
from (\ref{phit}),
this result is important because it can help us 
in resolving discrete ambiguities \cite{GQ}.
In the SM, once $\sin 2 \beta$ is measured, we know $\beta$ with no 
ambiguity from the bounds on the sides of the unitarity triangle.
However, in the presence of physics beyond the SM the values of the
``would be'' $\beta$ extracted from asymmetry measurements may
not fall within its SM allowed range. Such new physics cannot be
detected if the values of the asymmetry (i.e., $\sin 2 \beta$) 
lie within the SM range.
By measuring the sign of $\cos 2\beta$
we are sensitive to yet another kind of new physics:
new CP violating contributions to the mixing
amplitude.

\section{Conclusions}

In this paper we argued that photon polarization information in 
exclusive weak radiative $B$ decay can be used to probe new 
physics effects. The SM predicts that the photons emitted in 
$\bar B (B)$ decays are almost purely left (right) handed. 
By measuring the photon polarization we may find a signal for 
right-handed component that could only be generated by new physics.
Moreover, since the linear polarization states are also CP eigenstates,
the time-dependent CP asymmetries in $B^0(t)$ decays
are clean. In the SM they measure 
$\sin 2 \beta$; by comparing to the CP asymmetries in 
$B \to \psi K_S$ decay, a possible new CP violating amplitude in 
the $b \to s \gamma$ decay (independent on its helicity) can be found.

We discuss two methods for determining the photon polarization by using
the chain $B\to V\gamma \to Ve^+e^-$.
In the first method, discussed also in \cite{Mel,KKLM},
the photon is off-shell and we need to use 
the corresponding direct decay
$B\to Ve^+ e^-$ in the region where 
the dilepton invariant mass is close to the threshold since there 
photon exchange dominant the decay.
The second method makes use of the Bethe-Heitler process where photon
collide with matter and produce a lepton pair.

Two other methods were proposed in the past that are also
sensitive to a right-handed
component in the radiative decay amplitude. In \cite{Atwood} it was 
shown that
a CP asymmetry in the radiative mode can be generated by right-handed
amplitude. In \cite{MaRe} the $\Lambda_b \to \Lambda \gamma$
decay was studied and it was shown that
the $\Lambda$ polarization 
is sensitive to the right-handed operator $\tilde O_7$.

We comment next on the experimental feasibility of these four 
methods.
Experimentally, each method requires a different analysis, and
thus at this stage we can only estimate their relative efficiency.
As a benchmark we compare the efficiency of
each of these measurement to the efficiency of the
$B \to K^* \gamma$ rate measurement.

First, consider the method based on direct 
$B \to K^* e^+e^-$ decay. Here, the major obstacle is statistics as the
rate of the $B\to Ve^+ e^-$ decay mode in the $q^2 < 1$ GeV$^2$
region is smaller by about a factor of $10^{-2}$ compared to that of the
radiative decay.
However, electron pairs produced through virtual photons
are most sensitive to the photon polarization (the corresponding
efficiency parameter $\xi=1$). Moreover, we could hope that the 
efficiency of the dileptonic mode will be higher than that of 
the radiative mode.
The reconstruction of the dilepton pair emitted near the $B$ decay
point should be straightforward, 
as the corresponding tracks are expected to be well separated.
In the lab frame, the maximum value of the opening angle between the
$e^+$ and $e^-$ momenta is $\tan\theta_{\rm max} =
\sqrt{q^2-(2m_e)^2}/|\vec q\,|$.
For example, at $q_{\rm max}^2=0.5$  GeV$^2$ the maximum value of this
angle is  $\theta_{\rm max} =15^\circ$.
Moreover, at hadron colliders, where
the electron pair can be used for triggering, we could hope to get much
higher efficiency in the semileptonic mode compared to the radiative mode. 
Thus, our rough estimates indicate that this decay 
has an efficiency of the order of 
few percent compared to the measurement of the $B \to K^* \gamma$ 
decay rate.

Next we look at the method using Bethe-Heitler lepton pairs.  
There are two major drawbacks here. First,
the fraction of the photons that are converted is 
typically of the order of few percent,
depending on the detailed matter content of the experimental apparatus.
Second, the sensitivity to the photon polarization is 
not maximal ($\xi < 1$). On the positive side, we expect that the 
sensitivity to this mode will be higher compared to that 
of the $B \to K^* \gamma$ (where the $\gamma$
is identified in the calorimeter) as the energy resolution of the
lepton pair is higher and there is less background.
The momenta of the conversion electrons produced in the inner layers
of the detector are measured very well at CLEO (and the same is
expected to be true at BaBar and Belle). 

It seems that this kind of measurement
is easier to be carried out at $e^+e^-$ machines, as there is
much less background. Yet, it is not impossible that also at
hadron machines, where the statistics is much larger, this measurement
can be done.
We may conclude from our rough estimates that the efficiency for 
the Bethe-Heitler method is also at the few percent level.

The obvious advantage of the method proposed in \cite{Atwood} is that 
no polarization information is required. 
However, there are a few factors which offset this advantage. First, 
it is the
fact that time-dependent measurements are necessary. Second,
only neutral $B$ decays can be used, while in the methods we 
suggested, also charged $B$ (and $\Lambda_b$) can be used. Third,
flavor tagging is needed.
Last, the final state has to be a CP eigenstate, which 
gives further suppression of the rate through the chain 
$K^{*0}\to K_S \pi^0$. The combined effect of the above factors is
a reduction in the efficiency of about a factor of a 100.
We can conclude that this method seems somewhat disfavored 
compared to the methods we described.

Last we estimate the amount of data needed to carry out the 
suggestion of
Ref. \cite{MaRe}. Since this method required $\Lambda_b$ baryons,
it is clear that
this measurement can be done only at hadron machines. Moreover, 
$\Lambda_b$
baryons are produced only about $10\%$ of the time, and in general
are harder to identify than $B$ mesons. 
On the other hand it is relatively easy to collect the 
polarization information 
as the $\Lambda$ decay provide it with high efficiency.
Again, this very rough estimate suggests that if the radiative
decays can be seen at hadron machines, the  efficiency
for the $\Lambda_b \to \Lambda \gamma$ decay with polarization 
information
is at the few percent level compared to the $B \to K^* \gamma$ rate.

Finally, we comment on the
feasibility of the CP asymmetries measurements
discussed in Section III.
It seems that these measurements
are harder to perform since
both polarization information and time dependent
measurements are
needed; thus they suffer from the problem of both our
methods and the method of Ref. \cite{Atwood}.
Yet, when we try to resolve discrete ambiguities only 
the sign of $\cos 2\beta$ is needed. Clearly, 
the sign of a specific quantity 
can be determined more easily than its magnitude, and requires less
data. Therefore, we could still hope that 
the large numbers of $B$ mesons expected to become available at
the hadronic machines would make such measurements feasible.

Clearly, only a detailed experimental analysis can see which 
method is realistic. According to our estimates, it is possible
that all the different analyses discussed above will
be carried out. 


\acknowledgements
We are grateful to Jo\~ao Silva for helpful comments and in 
particular for 
pointing out to us the sensitivity of the time-dependent asymmetry 
to $\cos 2\beta$. We thank Helen Quinn for helpful discussions and
Stephane Plaszczynski, Soeren Prell,
Vivek Sharma and Abner Soffer
for useful discussions of the experimental aspects of the
methods presented here. 
Y. G. is supported by the U.S. Department of Energy under
contract DE-AC03-76SF00515.
The research of D. P. is supported in part by 
the DOE and by a National Science Foundation Grant no. PHY-9457911.


\begin{figure}[hhh]
 \begin{center}
 \mbox{\epsfig{file=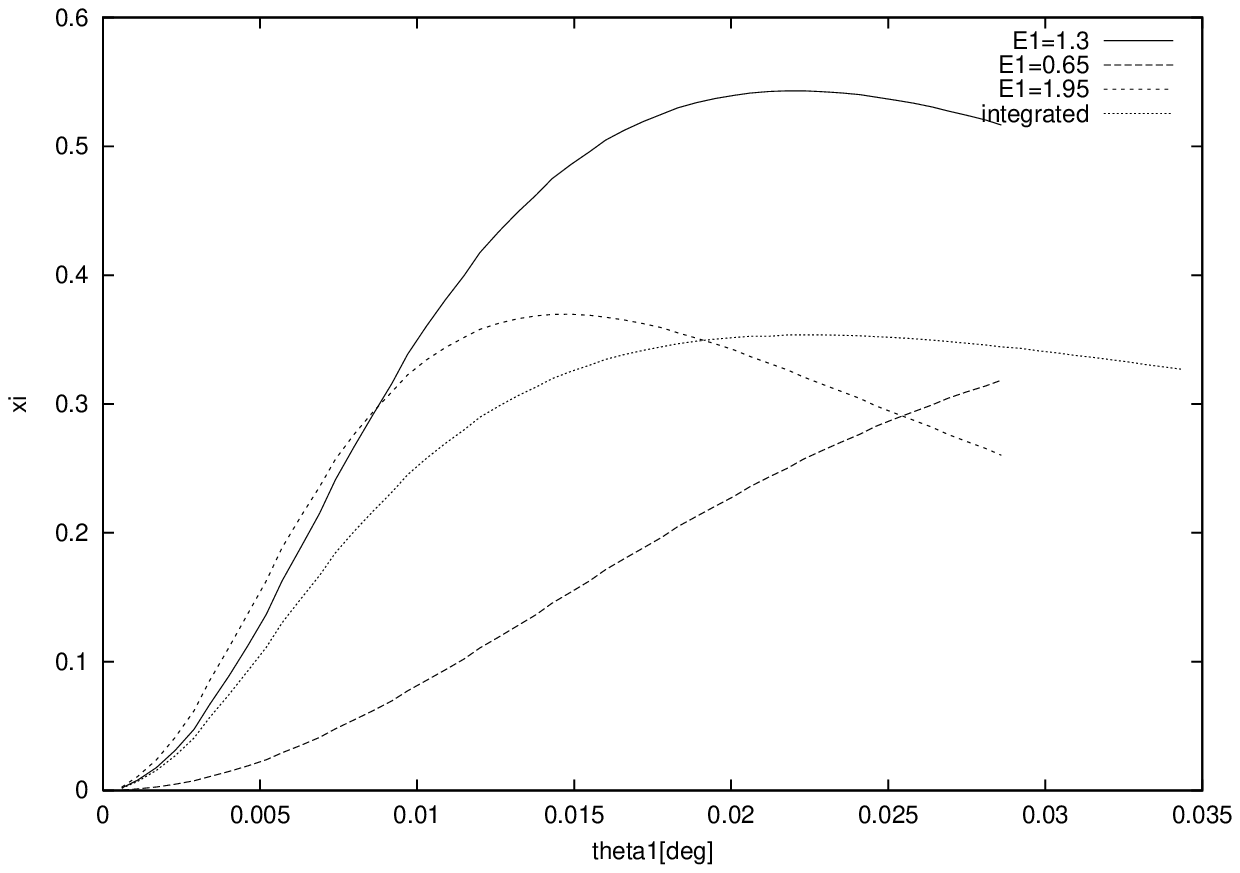,width=10cm}}
 \end{center}
 \caption{
The efficiency parameter $\xi$ appearing in the angular distribution
(\ref{angdist})
of the Bethe-Heitler pairs for a polarized $E_\gamma=2.6$ GeV photon.
The solid and dashed lines show the parameter $\xi$ for given positron
energies $E_1=1.3, 0.65$ and $1.95$ GeV, respectively. The dotted line
shows the $\xi$ parameter corresponding to an unobserved positron
energy.}
\label{fig1}
\end{figure}

\end{document}